# Machine Learning-Assisted Stability Boundary Determination of Multiport Autonomous Reconfigurable Solar Power Plants

Qianxue Xia, *Member, IEEE*, Suman Debnath, *Senior Member, IEEE, and* Maryam Saeedifard, *Fellow, IEEE*

***Abstract* — The multiport autonomous reconfigurable solar power plant (MARS) is a promising solution to integrate renewable energy resources and energy storage systems into the ac power grid and HVdc links. In the MARS system, various input power sources are connected to the individual submodules (SMs) through dc-dc converters. However, the presence of external power sources can result in unbalanced capacitor voltages of SMs, thereby violating stability constraints under multiple/diverse operating conditions. This paper aims to address the research gap by accurately determining the stability boundary for the MARS system. A novel machine learning (ML)-assisted energy balancing control (EBC) criterion is proposed. In conjunction with a refined EBC, this approach ensures balanced capacitor voltages across various types of SMs, significantly enhancing the overall system efficiency. The proposed EBC criterion effectively controls EBC activation and deactivation, achieving remarkable accuracy. Both PSCAD/EMTDC simulations and control hardware-in-the-loop (cHIL) tests are conducted to validate the feasibility and efficiency of the proposed method. By combining the EBC and ML-assisted EBC criteria, efficient energy management becomes achievable for systems featuring multiple input power sources, such as MARS. This approach enables the system to fully exploit its potential across an expanded operational range while upholding high efficiency standards.**

## I. INTRODUCTION

With the proliferation of photovoltaic (PV) generation in the power systems, the challenges posed by the intermittent nature of PV power and the decline in system inertia have become more pronounced. To address these challenges, hybrid solutions that integrate PV and energy storage systems (ESSs) have been explored and studied [1][2]. One of the proposed solutions is based on an integrated power converter connecting PV and ESS to high-voltage direct current (HVdc) links and ac power grids. In this context, an asymmetric modular converter, with PV submodules (SMs) in the upper arm and ESS SMs in the lower arm, is proposed in [3]. A similar topology for integrating photovoltaic (PV) systems and electric vehicle (EV) chargers is proposed in [4]. However, due to presence of high circulating current and power losses, such configuration is not suitable for high-power systems.

The concept of the multi-port autonomous reconfigurable solar power plant (MARS) has also emerged as a solution [5]. This system incorporates normal SMs, PV SMs, and ESS SMs, interconnecting utility-scale PV and ESS with HVdc links and ac transmission networks. The MARS, being modular and inherently reconfigurable, offers advantages over discrete hybrid PV-ESS plant development and HVdc systems, leading to reduced cost, enhanced efficiency, and improved reliability. In addition, the MARS can contribute to grid frequency support and boost the system inertia.

In conventional modular converter topologies, balanced capacitor voltages are usually maintained using sorting algorithms and similar methods [6]. However, alternative strategies, including Y-matrix modulation [7] and individual capacitor voltage balancing method [8], are also exist. However, these existing capacitor voltage balancing algorithms possess operational limitations and cannot ensure proper operation of the system when multiple types of SMs are present, leading to uneven power distribution in each arm of the converter.

Several papers have explored the operation and control of PV integration into the grid using modular topologies. Notably, some of these efforts [9]-[12] focus on addressing inter-submodule active power disparities to balance SM capacitor voltages. Reference [9] employs a hybrid MMC for integrating PV systems, wherein the full-bridge submodule offers increased flexibility for addressing power mismatches. Approaches in [10]-[11] employ individual control for each SM to counteract inter-SM power discrepancies caused by partial shading for medium voltage applications. However, the complexity of this strategy renders it unsuitable for modular topologies with a high number of SMs. Another strategy presented in [12] seeks to eliminate power mismatches within the modular topology under the conditions of partial shading of PV units or faulty cells by employing both dc and ac circulating currents. Nevertheless, this approach does not address the inter-SM power mismatch.

The presence of both normal SMs and ESS SMs within a modular converter-based ESS system is well studied in the literature [13]-[20]. In such a system, various factors, such as faulty battery units, recycled units, or redundant SMs, can lead to the presence of normal SMs. This diversity of SM types in the same arm can result in uneven distribution of intra-arms active power under specific operational conditions. The

Research sponsored by the Laboratory Directed Research and Development Program of Oak Ridge National Laboratory, managed by UT-Battelle, LLC, for the U. S. Department of Energy. Qianxue Xia, was with Georgia Institute of Technology, GA 30332 USA. She is now with Oak Ridge National Laboratory, Knoxville, TN 37932 USA (email: xiaq@ornl.gov).

Suman Debnath is with Oak Ridge National Laboratory, Knoxville, TN 37932, USA (email: debnaths@ornl.gov). Maryam Saeedifard is with the Electrical Engineering Department, Georgia Institute of Technology, GA, 30332 USA email: maryam@ece.gatech.edu).

operational limits of modular converter-based ESS systems have been studied in [14]-[20]. A PQ plot obtained from a power balance test is proposed in [14] to identify the stable operating range. Similarly, [16] suggests a criteria vector approach to determine stable operating conditions. However, methods introduced in [14]-[16] are conservative due to assumptions about the speed of sorting process and sorting frequency. Moreover, these methods are not applicable to the systems like MARS with more than two types of SMs within each arm. When there are more than two types of SMs, the dimension of unknown variables increases while the number of equations remains the same. The analysis in [17]-[20] assumes that the modular converter can be decoupled into ac and dc equivalent circuits, which does not apply to systems using the nearest level control (NLC) such as the MARS. The impact of capacitor voltage balancing control (CVBC) on system stability is significant, yet none of the cited references consider the CVBC algorithm when imposing the operating condition limits. Additionally, their approaches for estimating the inter-SM active power disparity limits are conservative, assuming ideal capacitor voltage balancing and unrestricted charge/discharge capability for each SM type.

MARS SMs fall into categories based on external power sources interfacing with the front-end half-bridge: normal SMs, PV SMs, and ESS SMs [5]. The number of PV SMs, ESS SMs, and the normal SMs in each arm depends on PV/ESS power rating and dc/ac voltages. Control and power management of hybrid PV-ESS modular converters are scarcely discussed in the literature [3], [21]. The asymmetric topology in [3], due to its significant power losses, is unsuitable for high-power PV systems. Reference [21] addresses a small-scale system with 9 SMs per arm, neglecting normal SMs and inter-SM power mismatch.

Machine learning (ML) is gaining traction in power electronics, primarily for system identification and control. In this context, neural networks are used to ease the computational load of model predictive control of a modular converter [24]-[26], enhance weight factor tuning [27], and estimate SM capacitor voltage [28]. Neural networks have even been utilized for circulating current reference calculation during unbalanced grid faults [29].

This paper introduces an innovative approach for evaluating the stability boundary of the MARS system, a complex structure with multiple types of input sources and a large array of SMs. The paper extends the operating region through a refined energy balancing control (EBC) method, addressing capacitor voltage imbalances and achieving intra-arm, intra-phase, and inter-phase power balance via a fundamental circulating current. The EBC criterion is formulated to minimize the circulating current, enhancing efficiency and mitigating current stress along with overall switching frequency. Different ML algorithms for the EBC criterion are developed and compared, applied to the MARS, and evaluated through control hardware-in-the-loop (cHIL) tests. This comprehensive investigation offers insights into improving the efficiency and reliability of the MARS and similar systems.

## II. System Modeling

### A. MARS System Topology

The circuit topology of the MARS system is shown in Fig. 1. The dc side of the MARS system is connected to an HVdc link while the ac side connects to the utility grid. The MARS configuration entails three phase-legs, wherein each leg has an upper and a lower arm. The symmetry extends to the upper and lower arms, with each arm accommodating a total of N SMs from three distinct types, the normal SM, the PV SM, and the ESS SM, as elaborated in the introduction section.

In scenarios where specific operational conditions prevail, variations in the capacitor voltages of different SMs can emerge. Stability is only upheld when the fluctuations in capacitor voltages remain confined within a predetermined range. To establish system stability, the divergence in SM capacitor voltages is managed and measured, defined by the following equation:

$$\sum(V_{c\sigma})^2 = \sum_{i=1}^{N}(v_{c,p,i,a} - \frac{v_{p,a}}{N})^2 + \sum_{i=1}^{N}(v_{c,n,i,a} - \frac{v_{n,a}}{N})^2 + \sum_{i=1}^{N}(v_{c,p,i,b} - \frac{v_{p,b}}{N})^2 + \sum_{i=1}^{N}(v_{c,n,i,b} - \frac{b}{N})^2 + \sum_{i=1}^{N}(v_{c,p,i,c} - \frac{v_{p,c}}{N})^2 + \sum_{i=1}^{N}(v_{c,n,i,c} - \frac{v_{n,c}}{N})^2 \quad (1)$$

where $v_{c,p,i,a}$ is the ith SM capacitor (c) voltage of the upper arm (p) for phase a. The cumulative sum $\sum(V_{c\sigma})^2$ characterizes the deviation of the capacitor voltages from their nominal values. The primary objective of control is to minimize $\sum(V_{c\sigma})^2$ to ensure that the capacitor voltages of different types of SMs do not diverge from each other. This guarantees intra-arm power balance within the system.

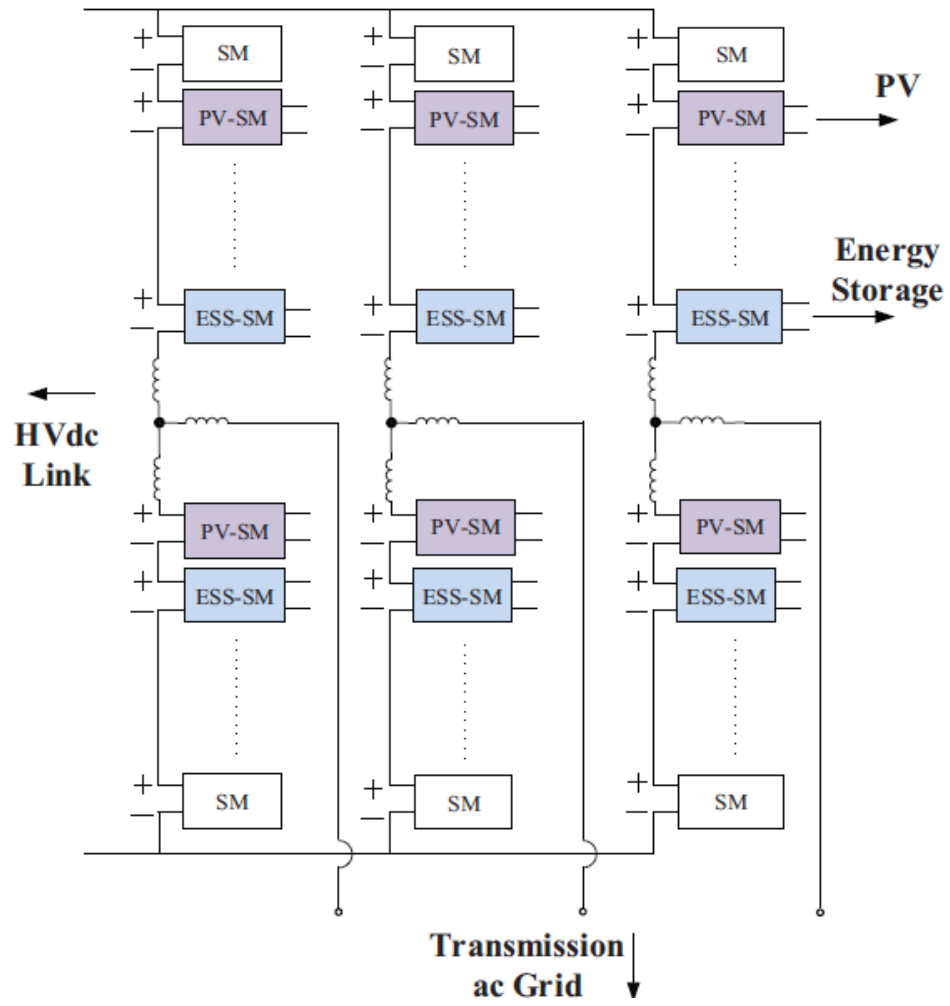

Fig. 1. MARS circuit architecture.

### B. MARS Hierarchical Control

The hierarchical control architecture of the MARS system includes three distinct levels denoted as L1, L2, and L3 and shown in Fig. 2. The L1 controller encompasses multiple functionalities, including virtual synchronous generator (VSG) control, PQ calculation, dq current control, EBC, and circulating current control. Positioned as the initial layer of control, L1 manages active and reactive power dispatch commands $P_{ac,ref}$, $Q_{ac,ref}$ transferred to the ac grid, along with the power transferred to the dc side, denoted as $P_{dc,ref}$. The EBC is implemented to balance the upper and lower arm SM capacitor voltages as well as the intra-arm capacitor voltages, by injecting a fundamental frequency circulating current. The reference for the injecting this circulating current is derived by minimizing $\sum(V_{c\sigma})^2$. The modulation index generated by the L1 controller is sent to the NLC, which forms the L2 controller along with the CVBC. The L2 controller aims at balancing the capacitor voltages of all different types of SMs, based on a capacitor voltage balancing algorithm influenced by [5]. This algorithm introduces upper and lower limit constraints on the capacitor voltages, leading to an increased efficiency by modifying the approach from turning on/off the first SM capacitor to turning on/off the capacitor with the lowest/highest voltage.

The L3 controller constitutes individual controllers for each PV and ESS SM, regulating power transfers from the PV and ESS. In the PV-SM L3 controller, a dual-loop control strategy is implemented. The outer-loop tracks the PV power reference stipulated by the L1 controller, while the inner-loop governs the inductor current of the dc-dc converter. Similarly, the ESS bi-directional dc-dc converter involves both an outer-loop power control and an inner-loop current control.

Incorporating switching states into the dynamics of the SM capacitor voltage difference introduces complexity to their formulation. Consequently, establishing the relationship between $\sum(V_{c\sigma})^2$and $P_{ac}$, $Q_{ac}$, $P_{dc}$ becomes intricate and time-consuming. To address this challenge, this paper introduces a data driven method to effectively ascertain the stability boundary, or in other words, the operating region of the MARS system.

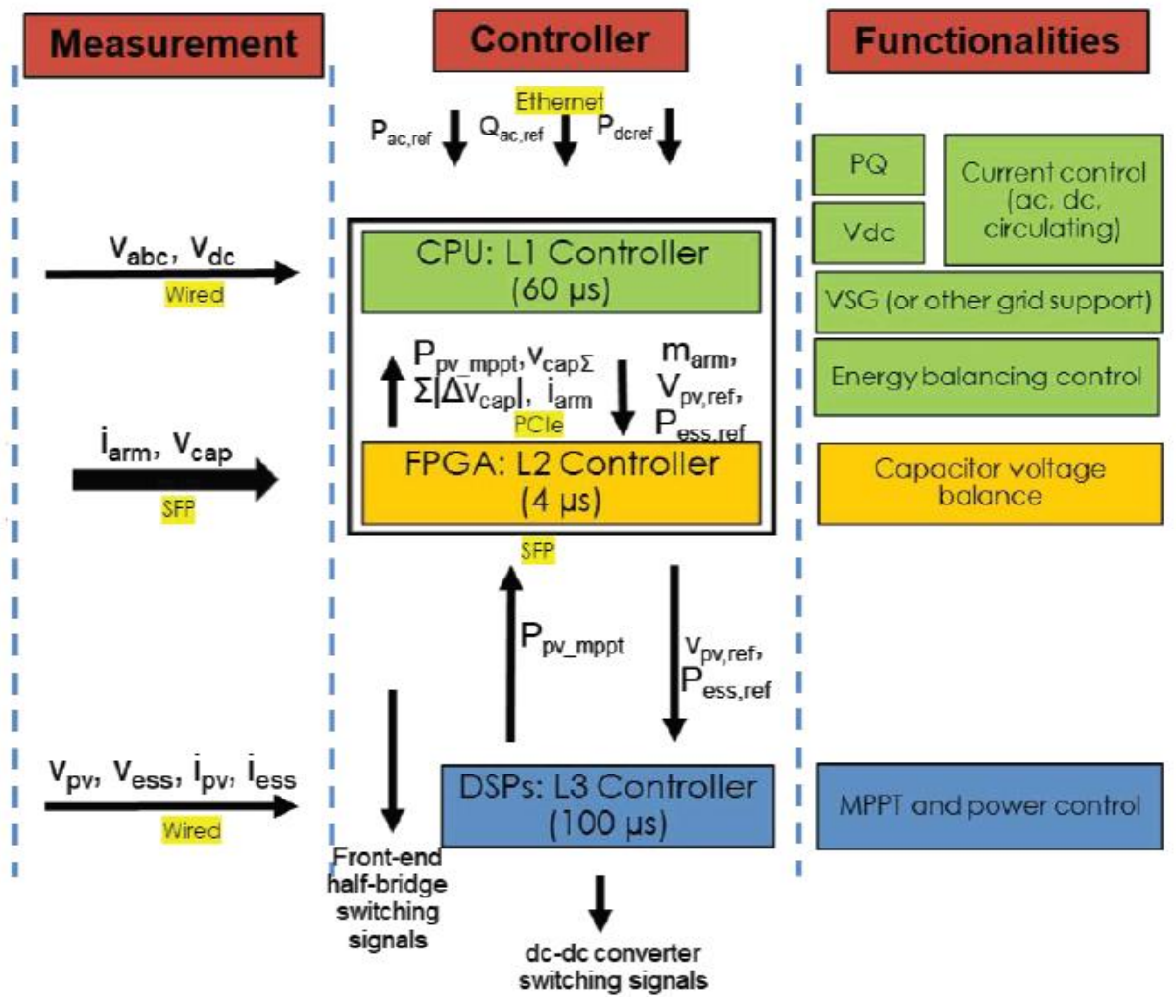

Fig. 2. Hierarchical control system to control the MARS.

## III. Capacitor Voltage Disparity

For the MARS system, the L2 controller faces limitations in ensuring that the injected power to each SM remains within the confines of the maximum power transfer limits across all operational scenarios. This challenge is particularly pronounced when dc and ac powers are small. Consequently, the peak of the arm current tends to be small, resulting in a limited energy storage capacity within the capacitors. To tackle this issue, the EBC is introduced to deal with the capacitor voltage disparity issue.

### *A. EBC Mathematical Model*

In the MARS system, the permissible ranges of SM capacitor voltages are established based on the designed stability constraints. The lower limit is determined to ensure stable operation of the L3 controller, while the upper limit is set to avoid the over-voltage stress on the semiconductor devices. Stability is deemed achieved when SM capacitor voltages remain within these limits. Apart from the voltage considerations, the dc current ripple is bounded to less than 20% of the nominal current. To confirm stability, both the capacitor voltage bounds and the dc current ripple constraint must simultaneously hold. Circulating currents offer a mechanism to distribute the external power within and between arms, boosting the charging/discharging capacity of the SMs without affecting the ac and dc sides of the MARS system. The EBC shown in Fig. 4 generates $P_{dj,ref}$ and $\sum Q_{dj,ref}$ (j=a,b,c), respectively, through two functions: (1) reducing the differences between the upper and lower arm voltages and (2) controlling $\sum(V_{c\sigma})^2$ to its reference value ($v_{c\delta ref}{}^2$). The circulating current references are then calculated based on the refined EBC, which enhances the approach in [13] and operates effectively without considering negative-sequence components. The fundamental circulating current references can be determined by:

$$i_{circ1q} = \frac{2v_{y,j}}{3}(P_{da,ref} + P_{db,ref} + P_{dc,ref}) \tag{2a}$$

$$i_{circ1d} = \frac{2v_{y,j}}{3}(Q_{da,ref} + Q_{db,ref} + Q_{dc,ref}). \tag{2b}$$

### *B. EBC Criterion*

In practical applications, the presence of circulating currents poses an operational challenge due to heightened current stress on system components, escalated magnitude of submodule capacitor voltage fluctuations, and increases conduction losses. Consequently, effective suppression of these currents is imperative for sustaining the overall performance and efficiency of the system. An advantageous approach would involve a criterion capable of enabling or disabling the EBC based on the system's inherent capacity to uphold intra-arm power balance. However, devising such an EBC criterion relies significantly on the stability boundary of the MARS system in the absence of EBC intervention. Comprehensive insights into the proposed EBC criterion are outlined in Section IV and visually represented Fig. 4.

### *C. Sorting Frequency of CVBC*

Apart from the intrinsic parameters such as the SM capacitance, which will directly impact the power charging/discharging capabilities, the frequency at which SM capacitor voltages are sorted holds significant sway over power disparity limits. This frequency's influence cannot be discounted when assessing the EBC criterion. Yet, the existing literature addressing the inherent operational constraints of modular converters often overlooks sorting frequency, often assuming it to be ideal. In a bid to bridge this gap, this section delves into the significance of sorting frequency.

Numerous SM capacitor voltage balancing algorithms have surfaced in the literature, with the aim of enhancing sorting efficiency — reducing computational load and simulation time — while maintaining a low switching frequency. To underscore the interplay between these algorithms and the system's operational region, a conventional capacitor voltage balancing algorithm [30] is subjected to simulation under varying sorting frequencies. In these simulations, the time step is set at h and sorting frequencies of h, 10h, 16h, and 20h are compared (ignoring the maximum switching frequency constraint of the switches). The correlation between the sorting frequency and the proportion of the stable operating conditions is shown in Fig. 3. Evidently, sorting frequency and the percentage of stable operating conditions exhibit a positive correlation. As a result, deriving the operational boundary mandates consideration of the chosen capacitor voltage balancing algorithm.

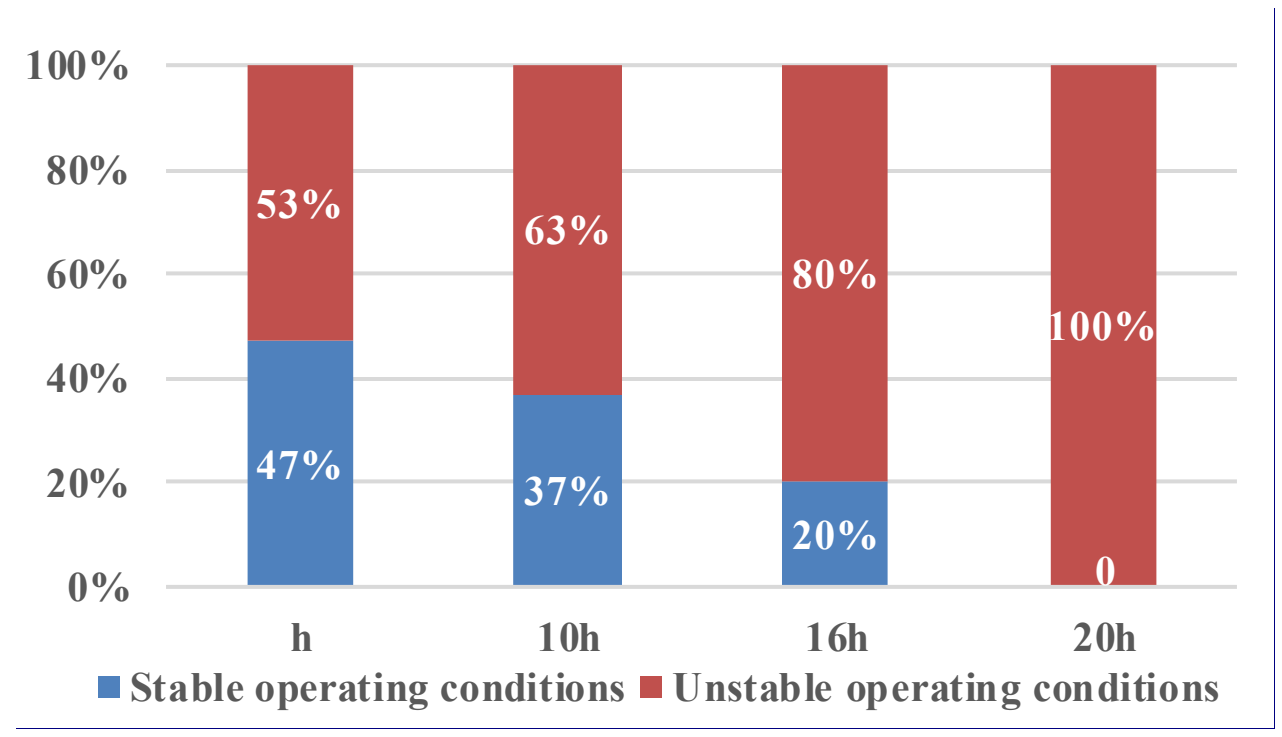

Fig. 3. Proportion of stable and unstable operating conditions vs the sorting frequency of the SM capacitor voltage balancing algorithm.

Analytical derivation of the sorting process is not trivial due to its nonlinearity. To circumvent complex analytical computations for mapping $(V_{c\sigma})^2$ to the EBC criterion, this paper employs ML algorithms such as Random Forest (RF) and Artificial Neural Network (ANN). These ML methods are harnessed to map the stability of operating conditions to input parameters like dc power, and ac active/reactive power.

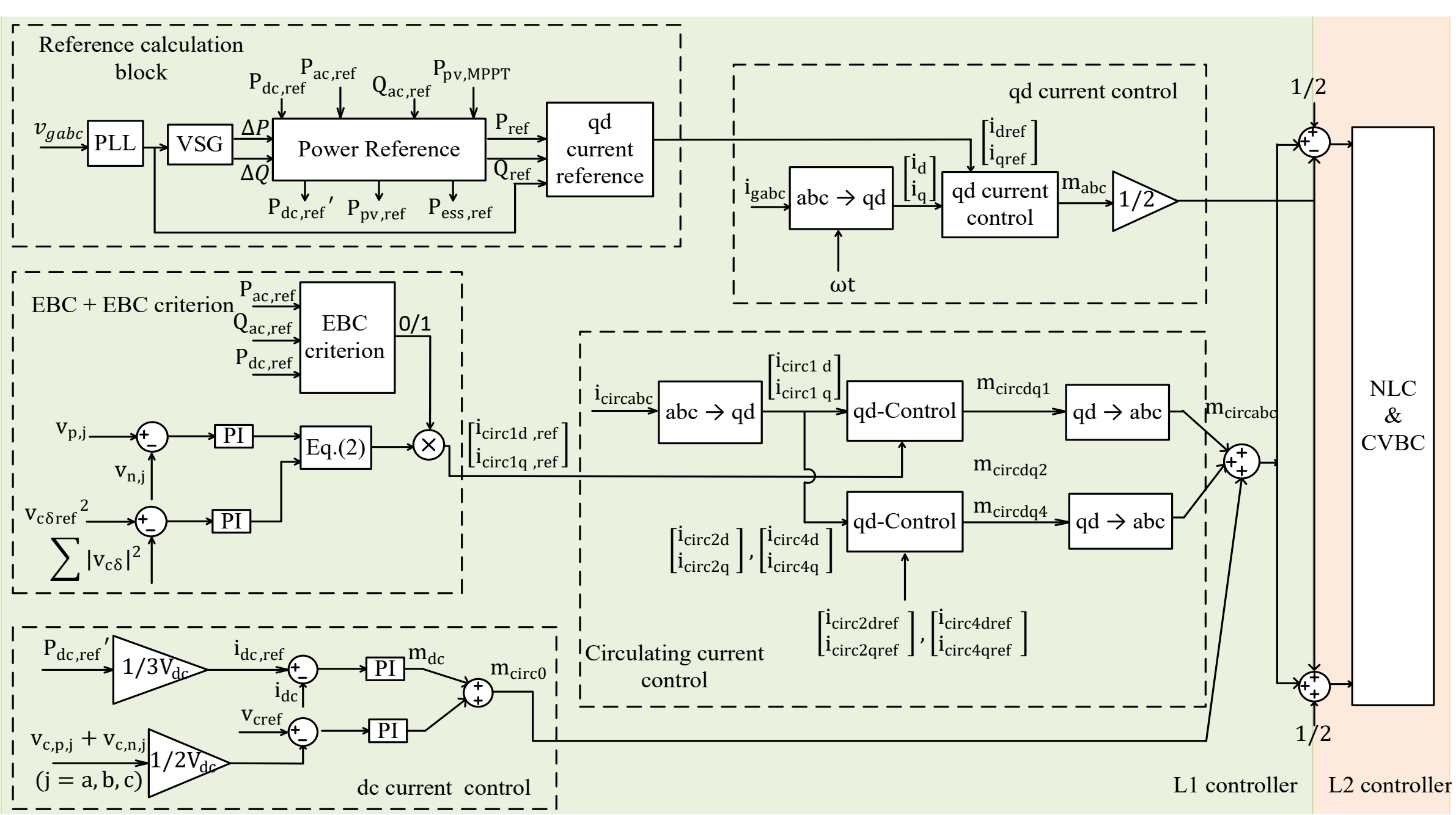

Fig. 4. High-level control architecture of the MARS (L1 and L2 controllers)

## IV. Learning-Assisted EBC Criterion

In order to implement the EBC criterion effectively, a non-linear classifier is essential. Hence, this paper opts for the utilization of two specific machine learning techniques: RF and ANN. These chosen methods serve as non-linear classifiers to facilitate the successful implementation of the EBC criterion.

### A. Data Acquisition

The data utilized in this study is extracted from the simulations conducted using the MARS high-fidelity model developed in [5]. This advanced model has proven to be significantly more efficient than the one based on the PSCAD/EMTDC library, boasting an 18,000 times faster simulation speed. This efficiency enables efficient data collection. Each simulation run, with a duration of 4 seconds, only takes approximately 8 minutes using the MARS system with 250 SMs per arm. Throughout the simulations, the input parameters $P_{ac}$, $P_{dc}$ and $Q_{ac}$ are swept from [-400, 400] MW, [-400, 400] MW with a resolution of 40 MW, and [-300, 300] MVar with a resolution of 30 MW, respectively. To ensure data accuracy, certain operating conditions are filtered out based on the constraints imposed by the $P_{pv_mppt}$, $P_{ess_rating}$ and power reference calculation block, as indicated in Fig. 4. The complete dataset encompasses a total of 6,158 operating conditions, which is further divided into three subsets: training (70%), validation (15%), and testing (15%). The training process for the neural network is conducted using the MATLAB neural pattern recognition toolbox, employing the scaled conjugate gradient back propagation algorithm. Once fine-tuned, the network is compiled into a MATLAB function and incorporated into the high-fidelity model in PSCAD through the MATLAB-PSCAD interface.

The stability of all 6,158 data points is effectively visualized in Fig. 5. Each operating condition is represented by a dot plotted with coordinates ($P_{ac}$, $P_{dc}$, $Q_{ac}$), where the color of the dot indicates the stability status of the condition—whether it is stable, voltage stable, or unstable—in the context of our system without the implementation of EBC and the EBC criterion. It becomes evident that a majority of operating conditions exhibiting $P_{ac}$ > 200 MW and $P_{dc}$ > 0 MW are stable. Positive $P_{dc}$ and highly positive $P_{ac}$ lead to a higher arm current, thereby enhancing the charging and discharging capacity of each SM capacitor and consequently improving the system stability. The term "voltage stable" shown in Fig. 5 refers to the operating conditions that satisfy the SM capacitor voltage constraints, even if their dc ripple magnitudes exceed the pre-specified upper bounds. Importantly, Fig. 5 illustrates the nonlinearity between the input and the output of the EBC criterion, affirming the rationale behind adopting ML algorithms for effectively mapping the input to the output.

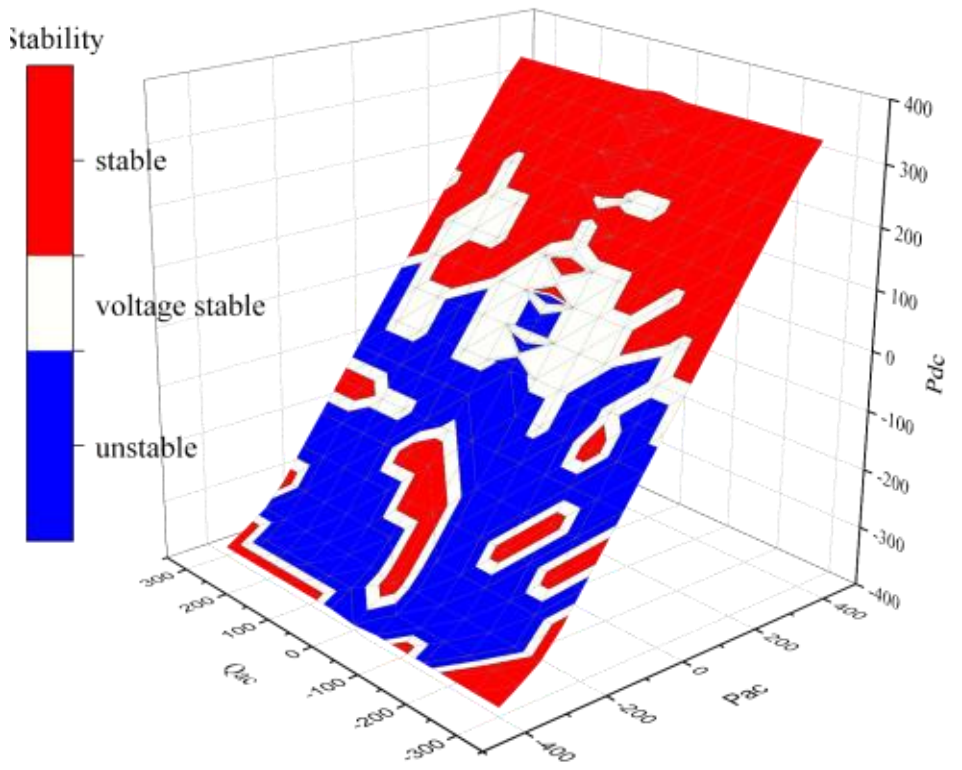

Fig. 5. Visualization of the stability for all 6,158 operating conditions.

### B. RF-Based EBC Criterion

The RF algorithm, which functions as an ensemble of decision trees, is leveraged for binary classification in this study. Its primary role is to make determinations regarding the necessity of the EBC. Each decision tree within the RF ensemble generates its own prediction for the class, either indicating EBC activation or deactivation. The final classification result is determined based on the class that receives the majority of votes from the decision trees in the RF ensemble. In essence, RF collectively aggregates the predictions from individual decision trees to reach an informed decision on whether the EBC should be turned on or off for a given operating condition.

### C. ANN-Based EBC Criterion

The design of the neural network-based controller operates under the premise of treating the entire system as a black box, without relying on the mathematical model of the system. This feature renders the neural network-based controller applicable even in cases where precise system knowledge is lacking, making it a competitive choice for intricate systems. Pattern recognition, defined as the process of identifying patterns and similarities within data through algorithmic techniques, is integral to this approach. In this context, ANNs are employed to facilitate pattern recognition for generating EBC enabling signals. The training of the ANN involves the optimization of the neuron weights to develop a learning model capable of accurately mapping the input variables to the output variables using the provided training data. Essentially, the network is trained to predict the future behaviors based on historical information. The construction of the model in this paper is refined through grid search hyperparameter tuning. It culminates in a 2-layer feedforward neural network, utilizing sigmoid activation functions, which are defined as:

$$S(x) = \frac{1}{1+e^{-x}} \tag{3}$$

and softmax activation functions defined as:

$$\sigma(z)_i = \frac{e^{z_i}}{\sum_{j=1}^{K} e^{z_i}} \quad for\ i = 1, \dots, K\ and\ z = (z_1, \dots, z_K) \in \mathbb{R}^K \tag{4}$$

Various configurations with different numbers of neurons in the hidden layer have been examined and compared. After conducting thorough evaluations, it has been determined that a configuration with nine neurons in the hidden layer yields optimal results. The trained neural network, showcasing this configuration, is depicted in Fig. 6.

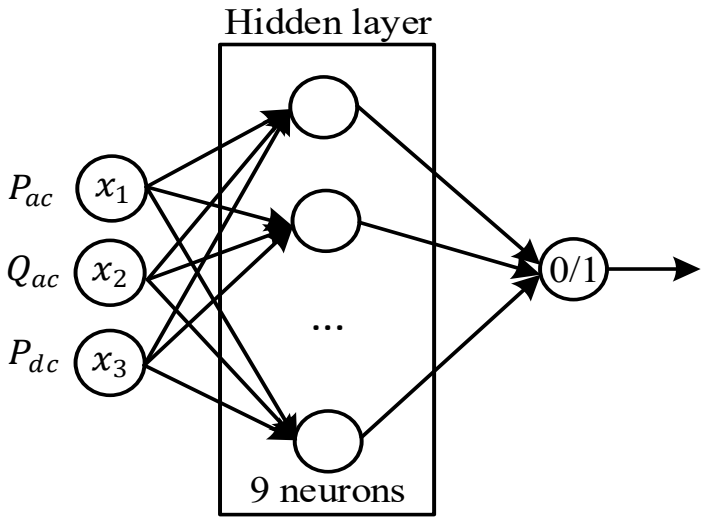

Fig. 6. Diagram of the ANN algorithm.

To assess the performance of the neural network classifier, confusion matrices have been generated and are presented in Fig. 7. The diagonal green cells represent the count of correctly classified cases, whereas the off-diagonal red cells represent cases that have been misclassified. Remarkably, the grey diagonal cells indicate exceptional recognition outcomes, achieving a training accuracy of 97.2%, a validation accuracy of 96.6%, and a testing accuracy of 96.5%.

### D. Comparison of Stability Boundary Determination Algorithms

Both the RF and ANN algorithms adopt an offline model training, resulting in a negligible computational burden during the inference phase. Consequently, the analysis in this section primarily focuses on the CPU utilization of the EBC criteria within the control system. This metric accurately reflects the practical computational demands of each method. The grid search method is employed for hyperparameter tuning for both RF and ANN algorithms. The CPU utilizations of the RF and ANN algorithms, as observed in control cHIL tests, are presented in Table 1. Notably, compared to the benchmark method without the EBC criterion, the increase in CPU utilization due to the deployment of RF and ANN algorithms is negligible and can be disregarded.

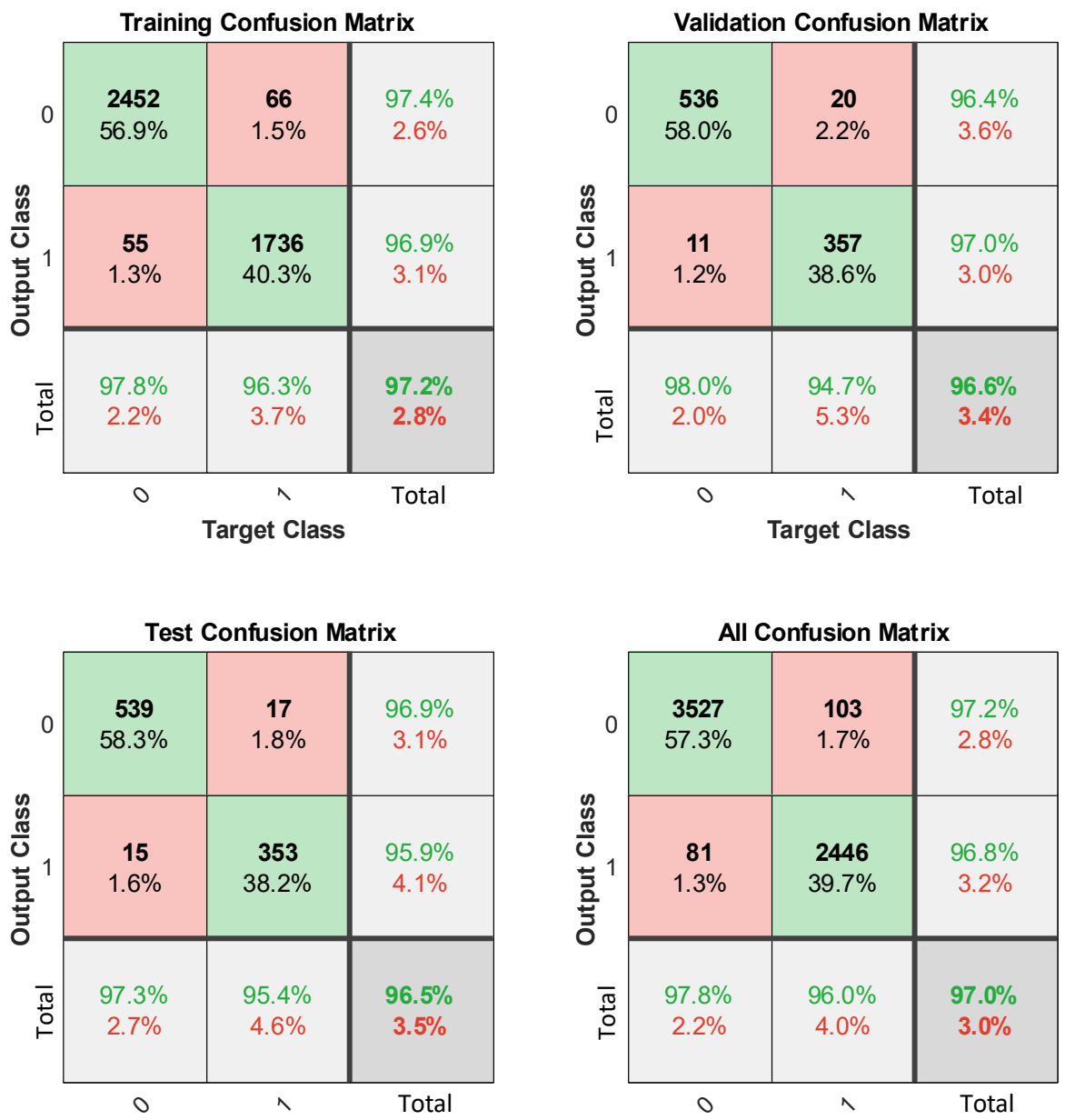

Fig. 7. Confusion matrix of the ANN training.

In Table 2, the benchmark method is demonstrated by a simple user-defined if-else statement, illustrated in Algorithm 1. Algorithm 1 determines the necessity for activating the EBC based on an analysis of system stability across 6,158 operating conditions without the EBC. From this analysis, the algorithm identifies the minimum $P_{ac}$ and $P_{dc}$ levels required for the system to remain stable. These minimum levels are set as $P_{ac_limit}$ and $P_{dc_limit}$, respectively. For any given operating condition, the algorithm checks if both $P_{ac}$ and $P_{dc}$ are above their respective limit values ( $P_{ac_limit}$ and $P_{dc_limit}$ ). If an operating condition satisfies these criteria ($P_{ac} > P_{ac_limit}$ and $P_{dc} > P_{dc_limit}$), it indicates that the system is stable without the need for EBC intervention. Therefore, in such scenarios, the EBC criterion disables the EBC by setting the references for the fundamental circulating current and its modulation index to zero.

The accuracy of the benchmark method is significantly lower than that of the RF and ANN methods. Furthermore, the overall accuracy of the ANN surpasses that of the RF algorithm. Consequently, for subsequent cHIL experiments and simulations in the PSCAD environment, the ANN algorithm is selected and tested due to its superior performance. The ANN model can be efficiently compiled using the MATLAB Simulink environment and subsequently integrated into the PSCAD high-fidelity and real-time experiment models. This selection ensures the use of an algorithm that provides enhanced accuracy and reliability in determining stability boundaries for the MARS system.

Algorithm 1: EBC criterion benchmark method

1. If $P_{ac} > P_{ac_limit}$ and $P_{dc} > P_{dc_limit}$
2. Set $i_{circq1ref} = 0$ and $i_{cird1ref} = 0$
3. Set $m_{cirq1} = 0$ and $m_{cird1} = 0$
4. end If

where $i_{circq1ref} = 0$, and $i_{cird1ref} = 0$ are the reference values of the fundamental circulating current; $m_{cirq1} = 0$ and $m_{cird1} = 0$ are the modulation indices of the fundamental circulating current.

Table 1 – CPU utilization in real time simulation

| | w/o EBC criterion | RF | ANN |
|---|---|---|---|
| CPU Utilization | 12.7% | 13.4% | 13.2% |

Table 2 – Testing accuracy of the confusion matrix

| | Benchmark method | RF | ANN |
|---|---|---|---|
| Accuracy | 80.57% | 96.05% | 96.5% |

## V. Simulation Results

The MARS study system, with its parameters listed in Table 3, represents an upgrade built upon the Trans Bay Cable project in Pittsburg, California [31]. The high-fidelity MARS model using the PSCAD/EMTDC software, as introduced in [5], is employed for conducting the simulations. The simulation framework includes the incorporation of a power loss calculation block to accurately assess power losses within the system. To ensure the simulation results are aligned with the corresponding cHIL results, additional elements such as communication delays have been thoughtfully integrated into the MARS model.

Table 3 – MARS system parameters.

| System Parameters | Value |
|---|---|
| Simulation time-step | 60 µs |
| System capacity (base power rating) | 400 MVA |
| System frequency | 60 Hz |
| Dc link voltage | ±200 kV |
| Ac terminal voltage | 220 kV |
| Arm inductor | 30 mH |
| Front-end half-bridge capacitor | 7.7 mF |
| Front-end half-bridge capacitor voltage | 1.6 kV |
| Number of PV-SMs/arm | 111 |
| Number of ESS-SMs/arm | 37 |
| Number of Normal-SMs/Arm | 102 |
| Total number of SMs/Arm | 250 |

In this study, simulations across 2500 randomly selected operating conditions were conducted to evaluate the system's losses both with and without the implementation of the Energy Balancing Control (EBC) criterion. Among these, for 1,396 operating conditions, the EBC criterion dictated that the EBC should be activated (output = 1). Under these conditions, there is no difference in power losses between scenarios with and without the EBC criterion, as the EBC would be activated in both cases.

Conversely, for the remaining 1,104 operating conditions, the EBC criterion suggested deactivation of the EBC (output = 0). In these scenarios, employing the EBC criterion leads to the EBC being deactivated, whereas, without the criterion, the EBC would remain activated across all conditions. The deactivation of the EBC under the criterion aims to reduce unnecessary activation of EBC and, consequently, minimize losses by injecting reduced fundamental circulating current.

To quantify the impact on losses, the relative loss reduction was calculated using the formula $(P_{loss}^{with\ EBC\ criterion} - P_{loss}^{without\ EBC\ criterion})/P_{loss}^{with\ EBC\ criterion}$. This formula represents the y-axis in Fig. 8, while the x-axis corresponds to each of the operating conditions evaluated.

The observed variability in loss reduction percentages can be attributed to the differing requirements for injected circulating current across operating conditions. A lower percentage indicates scenarios where the circulating current needed for capacitor voltage balancing was minimal. Conversely, a reduction nearing 50% signifies conditions that demanded significantly higher circulating currents for voltage balance.

This analysis highlights the EBC criterion's effectiveness in discerning when the activation of EBC is beneficial for optimizing system efficiency.

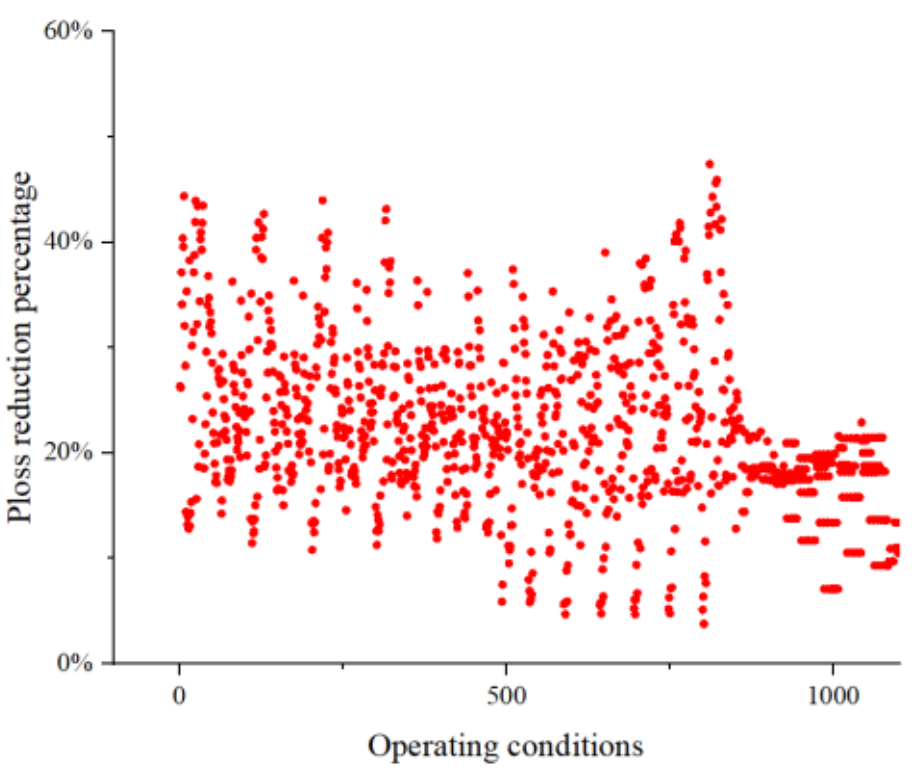

Fig. 8. Power loss reduction percentage for 1,104 operating conditions.

When the EBC criterion is employed, operating conditions that lie within the operational boundary exhibit reduced injected circulating current. This reduction in circulating current, in turn, leads to lower power losses within the system. Table 4 illustrates a collection of key statistics related to power loss reduction resulting from the EBC criterion's application ($P_{loss}^{with\ EBC\ criterion}$-$P_{loss}^{without\ EBC\ criterion}$). The data underscores the significance of the EBC criterion in terms of power savings. Across 1,104 test cases, the average reduction in power loss amounts to 550 kW. This substantial reduction in power losses substantiates the critical role played by the EBC criterion in enhancing system efficiency.

Table 4 – Statistics of the data w. / w.o. the EBC criterion

| | Power losses reduction (W) |
|---|---|
| Average | 550,319 |
| Median | 489,870 |
| Maximum | 1,582,270 |
| Minimum | 91,415 |
| Standard derivation | 306,060 |

## VI. cHIL Test Results

### A. cHIL System Setup

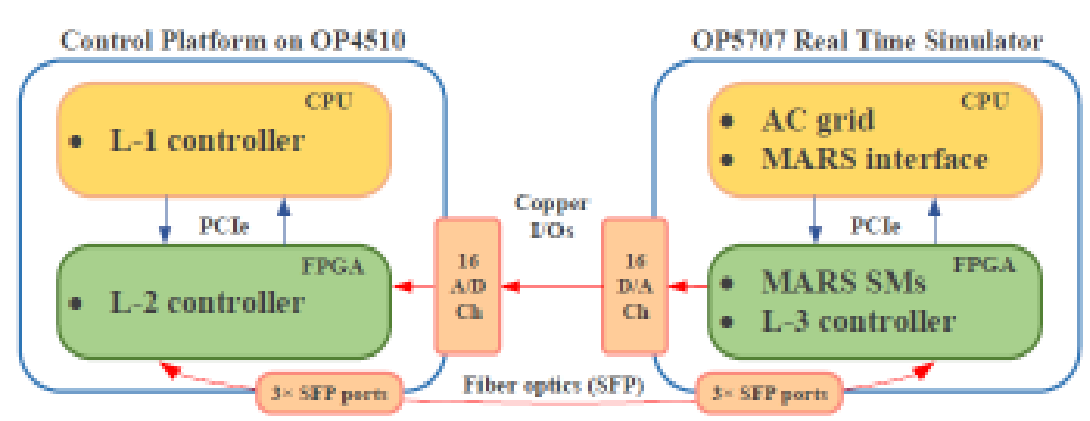

Fig. 9. Diagram of the cHIL system setup.

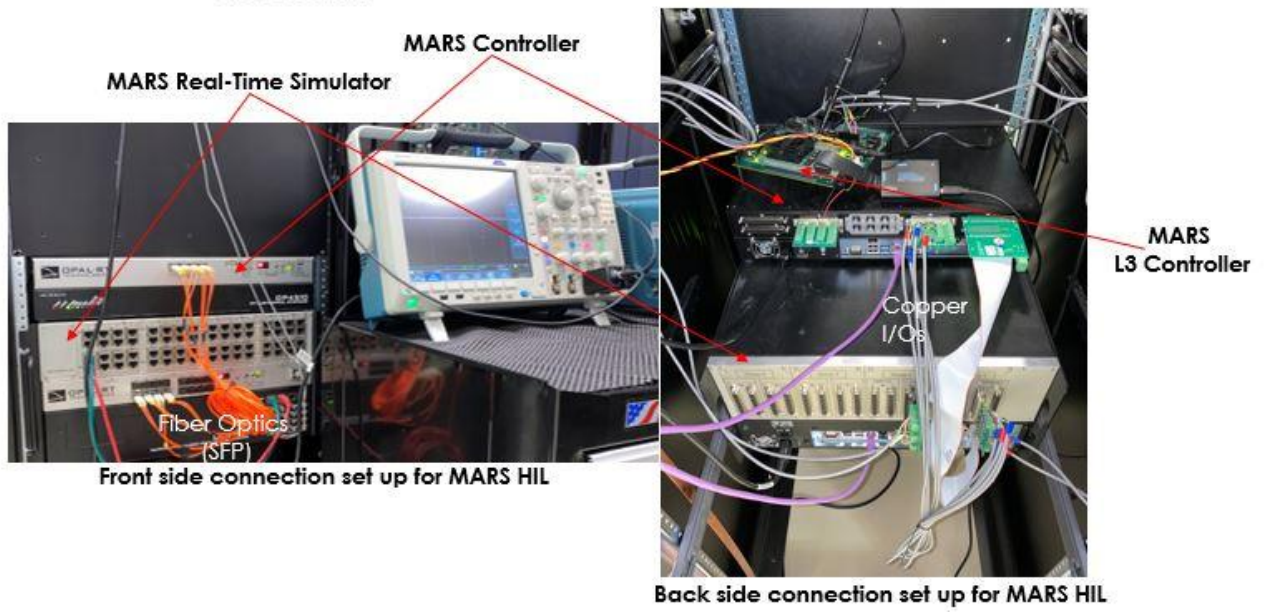

Fig. 10. The MARS cHIL test layout.

The MARS cHIL platform is constructed through the integration of the MARS control system connected to a real-time simulator. As shown in Fig. 9, the MARS converter, including the front-end half-bridge converter, along with the normal, PV, and ESS SMs, and their L3 controllers, are implemented in the Xilinx Virtex-7 field-programmable gate array (FPGA) of OPAL-RT's OP5707 platform while the ac grid model is implemented in the Central Processing Unit (CPU) of OP5707. This forms the real-time simulator. The ac grid model of MARS is based on the high-fidelity electromagnetic transient (EMT) model of the grid developed in [32]. The MARS control platform consists of L2 controller implemented on the FPGA and the L1 controller including the proposed EBC and EBC criterion in the CPU. A large number of signals need to be exchanged between the real-time simulator and the control platform. The small form-factor pluggable (SFP) communication medium is used to transfer the voltage and power references of the L3 controller from L1 controller in the control platform and map the data between the real-time simulation of the dynamics of MARS hardware and the control platform. To send the grid measurement from the real-time simulator to the control platform, the copper I/Os are utilized to replicate the real-world communication [33]. The different modes of data exchange are designed according to the signal size and latencies. The real-time simulator and the control platform, as well as the communication cables, are displayed in Fig. 10.

The presented results are per unitized with the system base power rating of 400 MW and the base L-L voltage of 220 kV. The base dc current is 1 kA and the base SM capacitor voltage of 1.6 kV.

### B. cHIL Test Results

#### 1) *Transition from Conventional Sorting Algorithm to EBC under Unstable Operating Conditions.*

This test case aims to demonstrate the limitations of the conventional sorting algorithm in balancing the capacitor voltages of the MARS system under certain operating conditions, and how the activation of the EBC addresses these limitations. Specifically, in the time frame from t = 500s to t = 514s, the MARS real-time simulation model operates without the EBC activated under the operating condition $P_{ac} = 0.25$ pu, $Q_{ac} = 0$, and $P_{dc} = 0.15$ pu. The PV and ESS dc-dc converters are enabled and connected to the MARS system with a PV reference voltage of 913 V and ESS reference discharging power of -147 kW.

Observations from Fig. 11 show that the capacitor voltages of the SM front-end half-bridge are initially unbalanced, deviating from the specified range of 0.85 pu to 1.15 pu (±15% of the front-end half-bridge capacitor's nominal voltage). This imbalance leads to instability in the L3 ESS controller. The EBC is subsequently activated around t = 514s, and the transition period of enabling the EBC causes some ripples between t = 514s and t = 519s. However, this ripple effect can be considered negligible since, in practical scenarios, the EBC would be activated initially under such operating conditions.

After t = 519s, as illustrated in Fig. 11, the activation of the EBC stabilizes the SM capacitor voltages, demonstrating the EBC's successful operation in ensuring SM capacitor voltage balance within the system.

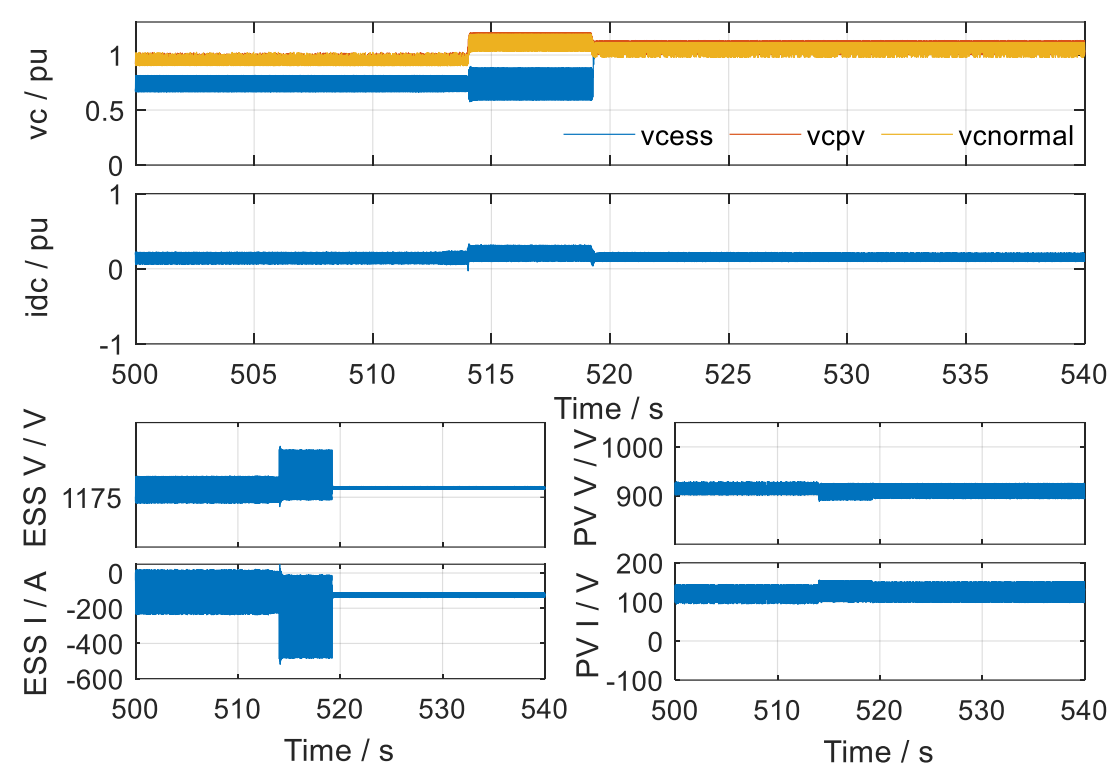

Fig. 11. Capacitor voltages in representative SMs subsequent to changes in the EBC criterion; ESS and PV side capacitor voltages and inductor current.

2) *Impact of Operating Condition Changes with EBC Criterion*

This section presents cyclic test results that assess the system's stability across varying steady-state operating conditions, managed by the integration of a NN-based EBC criterion within the control architecture. Fig. 12 illustrates the system's response to a series of dispatched command steps, with the EBC dynamically enabled or disabled based on the EBC criterion's output.

The test sequence initiates in a stable operating condition where $P_{ac} = 0.75$ pu, $Q_{ac} = 0$, and $P_{dc} = 0.75$ pu, which is a stable operating condition without EBC. The sequence then transitions to various operating conditions, all of which would be unstable without EBC intervention: first to $P_{ac} = 0.25$ pu, $Q_{ac} = 0$ and $P_{dc} = 0.15$ pu; then to $P_{ac} = -0.1875$ pu, $Q_{ac} = -0.125$ pu, and $P_{dc} = -0.185$ pu; followed by $P_{ac} = -0.625$ pu, $Q_{ac} = 0$ $P_{dc} = -0.625$ pu; and finally to $P_{ac} = -0.125$ pu, $Q_{ac} = 0$, and $P_{dc} = -0.185$ pu, before returning to the initial conditions. The simulation results shown in Fig. 12 indicate that the system remains stable throughout the changes in the operating conditions. The NN-based EBC criterion successfully generate the correct command in response to the varying conditions Furthermore, the transitions between different states are swift and exhibit acceptable overshoot. Notably, the capacitor voltages of the SM front-end half-bridges are maintained within the specified range, showcasing the successful collaboration of the hierarchical controllers and the EBC criterion. It also demonstrates the system's robust performance across a broad spectrum of operating conditions.

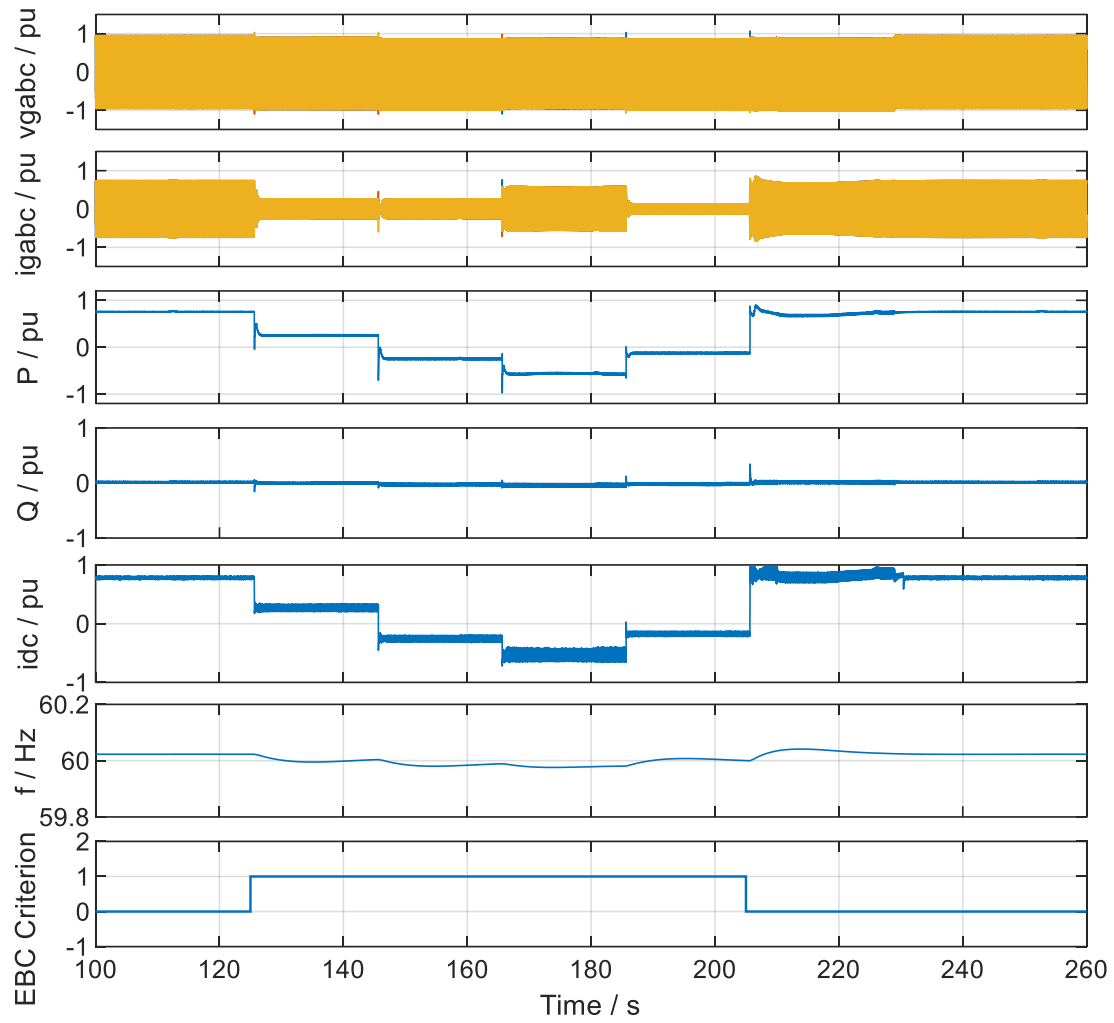

Fig. 12. MARS grid-side measured three-phase voltages and currents, active and reactive powers, dc current, grid frequency, and output of EBC criterion.

3) *Influence of $P_{dc}$ and $Q_{ac}$ on Capacitor Voltage Disparity*

These two test case is to test the influence of Pdc and Qac on SM capacitor voltage disparity, thus the system is operated without the EBC nor EBC criterion. This section presents two test cases designed to evaluate the influence of $P_{dc}$ and $Q_{ac}$ on the voltage disparity across SM capacitors. During these tests, the system operates without the EBC or the EBC criterion to isolate the effects of $P_{dc}$ and $Q_{ac}$.

As discussed in Section III, $P_{dc}$ is an important factor that impacts the capacitor voltage disparity. Since $i_{dc}$ is proportional to $P_{dc}$, larger values of $P_{dc}$ lead to higher arm currents, which subsequently impacts the higher charging/discharging rates of the SM capacitors. This effect is illustrated in Fig. 13, where the system starts in an operating condition of $P_{ac} = 0.25$ pu, $P_{dc} = 0.15$ pu, and $Q_{ac} = 0$. Under this operating condition, the capacitor voltages across different types of SMs are initially unbalanced, as shown in Fig. 13 from t = 1664 to 1682s. Upon increasing $P_{dc}$ to 0.2 pu at t = 1682s and subsequently to 0.25 pu at t = 1697s, the deviation in ESS SM capacitor voltage from other types of SM voltages gradually reduces, indicating that under certain conditions, a larger $P_{dc}$ can aid in balancing SM capacitor voltages. However, this relationship is complex and not directly proportional, as $P_{dc}$ is dependent on both the $P_{pv}$ and $P_{ess}$, and the combinations of $P_{pv}$ and $P_{ess}$ can also impact the SM capacitor voltages.

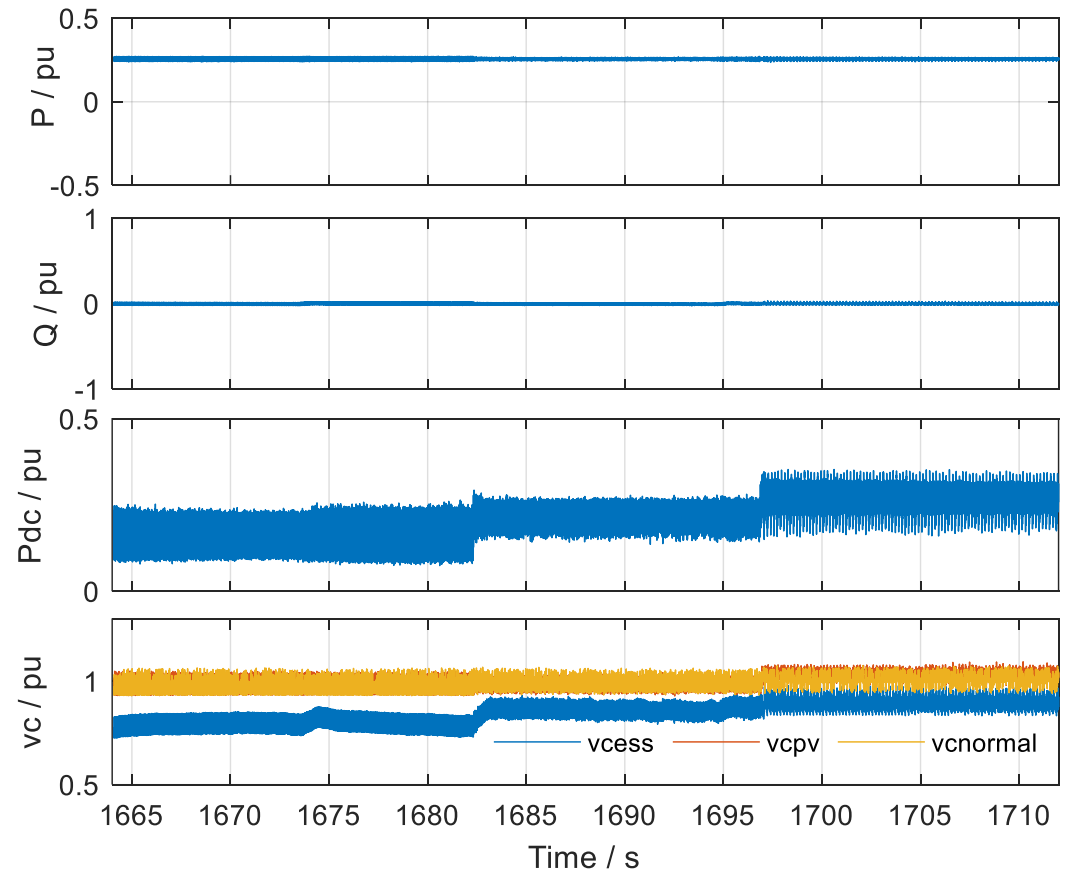

Fig. 13. MARS grid-side measured active and reactive powers, dc power, and capacitor voltages of different types of SMs.

Furthermore, increasing reactive power $Q_{ac}$ has been suggested as a method to stabilize unstable operating conditions, as stated in Reference [13]. The impact of $Q_{ac}$ on SM capacitor voltage disparity is examined initially under the operating condition of $P_{ac} = 0.25$ pu, $P_{dc} = 0.15$ pu, and $Q_{ac} = 0$, with results shown in Fig. 14. The system is subjected to a constant $P_{ac} = 0.25$ pu with $Q_{ac}$ stepping up from 0 to 0.5 pu at t = 1482s and subsequently stepping down to -0.5 pu at t= 1529s. The results in Fig. 14 show that the SM capacitor voltage balancing, i.e., intra-arm power balance, improves with none-zero $Q_{ac}$. Different operating conditions are tested in which $Q_{ac}$ has a positive effect on reducing the disparity between SM capacitor voltages. However, there are also some cases where this relationship does not hold true.

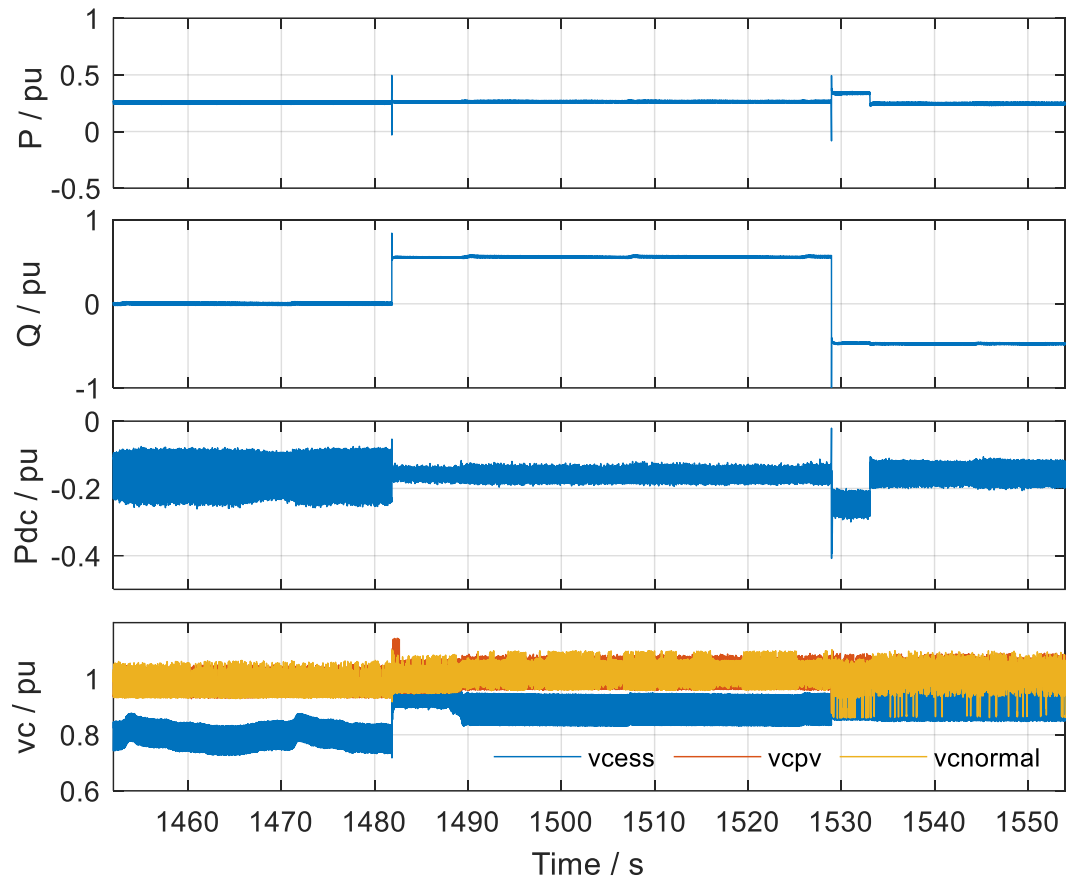

Fig. 14. MARS grid-side measured active and reactive powers, dc power, and capacitor voltages of different types of SMs.

## VII. Conclusions

This paper addresses the challenge of the SM capacitor voltage disparity in modular converter systems, which can arise due to limitations in the capacitor voltage balancing algorithm, low levels of direct current (dc) power, and low reactive power. This issue is tackled by introducing the EBC, which injects a fundamental circulating current to balance the SM capacitor voltages. However, applying the EBC can lead to additional power losses in scenarios where it might not be necessary.

To determine the boundary between stable and unstable operating conditions and to decide whether to activate the EBC, the paper utilizes two ML algorithms: ANN and RF. These algorithms offer a practical way to achieve high accuracy in determining the need for EBC, eliminating the need for complex analytical derivations. The ANN algorithm outperforms RF and is implemented in both simulation software (PSCAD) and a cHIL platform.

The results of the PSCAD simulations demonstrate the potential for power reduction across various operating conditions using the EBC criterion. The cHIL tests further validate the effectiveness of the EBC in addressing capacitor voltage disparities under certain conditions. These tests also explore the relationships between system stability, reactive power, and dc power. The cyclic tests conducted in the cHIL platform showcase the stable operation of the system across different power references while employing the EBC and its associated criterion.

The EBC method, complemented by the EBC criterion, is applicable to modular converter systems of any size that integrate SMs with various external power sources such as ESS and PV. The utilization of the EBC potentially extends the operating region of the system, and the incorporation of the EBC criterion can improve the efficiency.

## VIII. References

[1] J. Wang, K. Sun, C. Xue, T. Liu and Y. Li, "Multi-Port DC-AC Converter with Differential Power Processing DC-DC Converter and Flexible Power Control for Battery ESS Integrated PV Systems," in *IEEE Transactions on Industrial Electronics*, vol. 69, no. 5, pp. 4879-4889, May 2022.

[2] S. Kang, J. Noh and J. -W. Park, "Active Distribution Management System Based on Smart Inverter Control of PV/ESS Integrated System," in *IEEE Transactions on Industrial Electronics*, vol. 69, no. 8, pp. 7994-8003, Aug. 2022.

[3] L. Zhang, Z. Zhang, J. Qin, D. Shi and Z. Wang, "Design and Performance Evaluation of the Modular Multilevel Converter (MMC)-based Grid-tied PV-Battery Conversion System," *IEEE Energy Conversion Congress and Exposition (ECCE)*, 2018, pp. 2649-2654.

[4] Y. -C. Su, H. -M. Li, P. -L. Chen and P. -T. Cheng, "Integration of PV Panels and EV Chargers on the Modular Multilevel Converter Based SST," in *IEEE Transactions on Industry Applications*, vol. 58, no. 5, pp. 6428-6437, Sept.-Oct. 2022.

[5] S. Debnath *et al*., "Renewable Integration in Hybrid AC/DC Systems Using a Multi-Port Autonomous Reconfigurable Solar Power Plant (MARS)," in *IEEE Transactions on Power Systems*, vol. 36, no. 1, pp. 603-612, Jan. 2021.

[6] Y. Lyu, Y. Zi and X. Li, "Improved Control of Capacitor Voltage Balancing in Modular Multilevel Converter Submodules," in *IEEE Access*, vol. 12, pp. 19510-19519, 2024.

[7] Y. Liu and F. Z. Peng, "A Modular Multilevel Converter With Self-Voltage Balancing Part II: Y-Matrix Modulation," in *IEEE Journal of Emerging and Selected Topics in Power Electronics*, vol. 8, no. 2, pp. 1126-1133, June 2020.

[8] L. Zhou, C. Chen, J. Xiong and K. Zhang, "A Robust Capacitor Voltage Balancing Method for CPS-PWM-Based Modular Multilevel Converters Accommodating Wide Power Range," in *IEEE Transactions on Power Electronics*, vol. 37, no. 12, pp. 14306-14316, Dec. 2022.

[9] T. S. Basu and S. Maiti, "A Hybrid Modular Multilevel Converter for Solar Power Integration," in IEEE Transactions on Industry Applications, vol. 55, no. 5, pp. 5166-5177, Sept.-Oct. 2019.

[10] S. Rivera, B. Wu, R. Lizana, S. Kouro, M. Perez and J. Rodriguez, "Modular Multilevel Converter for Large-Scale Multistring Photovoltaic Energy Conversion System," *IEEE Energy Conversion Congress and Exposition*, 2013, pp. 1941-1946.

[11] F. Rong, X. Gong and S. Huang, "A Novel Grid-Connected PV System Based on MMC to Get the Maximum Power Under Partial Shading Conditions," in *IEEE Transactions on Power Electronics*, vol. 32, no. 6, pp. 4320-4333, June 2017.

[12] H. Bayat and A. Yazdani, "A Power Mismatch Elimination Strategy for an MMC-Based Photovoltaic System," in *IEEE Transactions on Energy Conversion*, vol. 33, no. 3, pp. 1519-1528, Sept. 2018.

[13] N. Li, F. Gao, T. Hao, Z. Ma and C. Zhang, "SOH Balancing Control Method for the MMC Battery Energy Storage System," in *IEEE Transactions on Industrial Electronics*, vol. 65, no. 8, pp. 6581-6591, Aug. 2018.

[14] T. Soong and P. W. Lehn, "Assessment of Fault Tolerance in Modular Multilevel Converters with Integrated Energy Storage," in *IEEE Transactions on Power Electronics*, vol. 31, no. 6, pp. 4085-4095, June 2016.

[15] A. Hillers and J. Biela, "Fault-Tolerant Operation of The Modular Multilevel Converter in an Energy Storage System Based on Split Batteries," *16th European Conference on Power Electronics and Applications*, 2014, pp. 1-8.

[16] G. Liang *et al*., "Analytical Derivation of Intersubmodule Active Power Disparity Limits in Modular Multilevel Converter-Based Battery Energy Storage Systems," in *IEEE Transactions on Power Electronics*, vol. 36, no. 3, pp. 2864-2874, March 2021.

[17] F. Briz, M. López, A. Zapico, A. Rodríguez and D. Díaz-Reigosa, "Operation and Control of MMCs Using Cells with Power Transfer Capability," *IEEE Applied Power Electronics Conference and Exposition (APEC)*, 2015, pp. 980-987.

[18] Y. Ma, H. Lin, X. Wang, Z. Wang and Z. Ze, "Analysis of Battery Fault Tolerance in Modular Multilevel Converter with Integrated Battery Energy Storage System," *20th European Conference on Power Electronics and Applications (ECCE Europe)*, 2018, pp. P.1-P.8.
[19] Z.Wang, H. Lin, and Y.Ma, "Improved Capacitor Voltage Balancing Control for Multimode Operation of Modular Multilevel Converter with Integrated Battery Energy Storage System," *IET Power Electron.*, vol. 12, no. 11, pp. 2751–2760, Sep. 2019.
[20] T. Soong and P. W. Lehn, "Internal Power Flow of a Modular Multilevel Converter with Distributed Energy Resources," in *IEEE Journal of Emerging and Selected Topics in Power Electronics*, vol. 2, no. 4, pp. 1127-1138, Dec. 2014.
[21] H. Bayat and A. Yazdani, "A Hybrid MMC-Based Photovoltaic and Battery Energy Storage System," in *IEEE Power and Energy Technology Systems Journal*, vol. 6, no. 1, pp. 32-40, March 2019.
[22] S. Rohner, S. Bernet, M. Hiller, and R. Sommer, "Modelling, simulation and analysis of a modular multilevel converter for medium voltage applications," in *Proc. IEEE Int. Conf. Ind. Technol*., 2010, pp. 775–782.
[23] S. Rohner, S. Bernet, M. Hiller, and R. Sommer, "Modulation, losses, and semiconductor requirements of modular multilevel converters," *IEEE Transactions on Industrial Electronics*, vol. 57, no. 8, pp. 2633–2642, Aug. 2010.
[24] S. Wang, T. Dragicevic, Y. Gao and R. Teodorescu, "Neural Network Based Model Predictive Controllers for Modular Multilevel Converters," in *IEEE Transactions on Energy Conversion*, vol. 36, no. 2, pp. 1562-1571, June 2021.
[25] X. Liu *et al*., "Predictor-Based Neural Network Finite-Set Predictive Control for Modular Multilevel Converter," in *IEEE Transactions on Industrial Electronics*, vol. 68, no. 11, pp. 11621-11627, Nov. 2021.
[26] S. Wang, T. Dragicevic, G. F. Gontijo, S. K. Chaudhary and R. Teodorescu, "Machine Learning Emulation of Model Predictive Control for Modular Multilevel Converters," in *IEEE Transactions on Industrial Electronics*, vol. 68, no. 11, pp. 11628-11634, Nov. 2021.
[27] X. Liu *et al*., "Neural Predictor-Based Low Switching Frequency FCS-MPC for MMC With Online Weighting Factors Tuning," in *IEEE Transactions on Power Electronics*, vol. 37, no. 4, pp. 4065-4079, April 2022.
[28] P. Poblete, G. Pizarro, G. Droguett, F. Núñez, P. D. Judge and J. Pereda, "Distributed Neural Network Observer for Submodule Capacitor Voltage Estimation in Modular Multilevel Converters," in *IEEE Transactions on Power Electronics*, vol. 37, no. 9, pp. 10306-10318, Sept. 2022.
[29] S. Wang, T. Dragicevic, Y. Gao, S. K. Chaudhary and R. Teodorescu, "Machine Learning Based Operating Region Extension of Modular Multilevel Converters Under Unbalanced Grid Faults," in *IEEE Transactions on Industrial Electronics*, vol. 68, no. 5, pp. 4554-4560, May 2021.
[30] S. Rohner, S. Bernet, M. Hiller, and R. Sommer, "Modulation, losses, and semiconductor requirements of modular multilevel converters," in *IEEE Transactions on Industrial Electronics*, vol. 57, no. 8, pp. 2633–2642, Aug. 2010.
[31] H. Knaak, "Modular Multilevel Converters and Hvdc/Facts: A Success Story," in *Proceedings of the 14th European Conference on Power Electronics and Applications*, Aug 2011, pp. 1–6
[32] S. Debnath, Q. Xia, M. Saeedifard, and M. Arifujjaman, "Advanced High-Fidelity Lumped EMT Grid Modelling & Comparison," *CIGRE Grid of the Future,* 2019.
[33] S. Debnath, P. R. V. Marthi, Z. Dong, Q. Xia and S. Chakraborty, "Power Electronic Hardware-in-the-Loop (PE-HIL): Testing Individual Controllers in Large-Scale Power Electronics Systems," *IEEE Energy Conversion Congress and Exposition (ECCE)*, 2022, pp. 1-8.

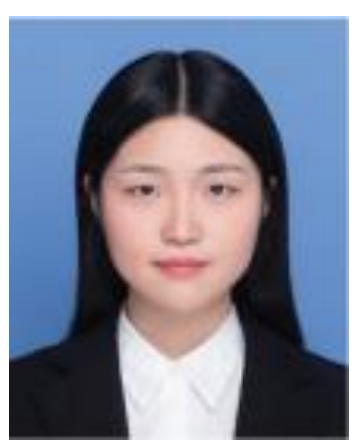

Qianxue Xia (Member, IEEE) She received the M.S. degree in electrical engineering from Arizona State University, Tempe, AZ, USA, in 2018. She received the Ph.D. degree in electrical engineering from in electrical engineering at Georgia Institute of Technology in 2022. Currently, she is a postdoctoral researcher at Oak Ridge National Laboratory, Knoxville, TN, USA. Her main research interests are in artificial intelligence application in control and modeling of power electronics.

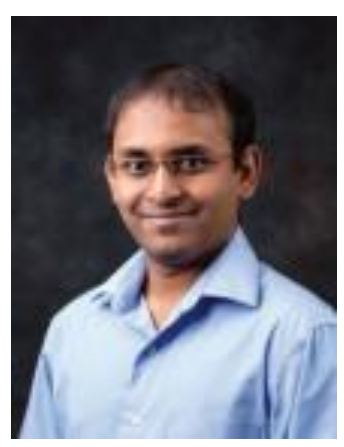

Suman Debnath (Senior member, IEEE) received the bachelors and masters degrees in electrical engineering from Indian Institute of Technology at Madras, Tamil Nadu, India, in 2010 and the Ph.D. degree in electrical engineering from Purdue University, West Lafayette, IN, USA, in 2015. Currently, he is a Senior R&D staff at Oak Ridge National Laboratory, Knoxville, TN, USA. His main research interests are in applied mathematics for high-fidelity modeling and advanced simulation algorithms for electromagnetic transient [EMT] simulation, among others.

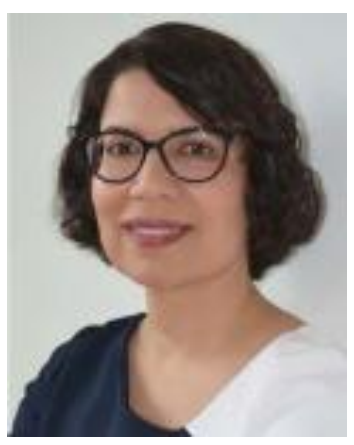

Maryam Saeedifard (Fellow, IEEE) received the Ph.D. degree in electrical engineering from the University of Toronto, Toronto, ON, Canada, in 2008. She is currently a Professor with the School of Electrical and Computer Engineering, Georgia Institute of Technology, Atlanta, GA, USA, where she holds a Dean's Professorship. Her research interests include power electronics and applications of power electronics in power systems.